\documentclass{iau} 
\usepackage{graphicx} \usepackage{rotating}
\usepackage{txfonts}
\usepackage{url} 
\usepackage{fmtcount}
\usepackage{upgreek}

\title[$^3$He Abundances in Planetary Nebulae] %% give here short title %%
{$^3$He Abundances in Planetary Nebulae}

\author[Lizette Guzman-Ramirez]   %% give here short author list %%
{Lizette Guzman-Ramirez$^{1,2}$\thanks{E-mail: guzmanl@strw.leidenuniv.nl}}

\affiliation{$^{1}$European Southern Observatory, Alonso de C\'ordova 3107, Casilla 19001, Santiago, Chile\\
$^{2}$Leiden Observatory, Leiden University, Niels Bohrweg 2, 2333 CA Leiden, The Netherlands\\}

\pubyear{2017}
\volume{323}  %% insert here IAU Symposium No.
\jname{Planetary Nebulae: Multi-wavelength Probes of Stellar and Galactic Evolution}
\editors{X. Liu, L. Stanghellini, \& A. Karakas, eds.}
\begin{document}

\maketitle

\begin{abstract}
The $^3$He isotope is important to many fields of astrophysics,
including stellar evolution, chemical evolution, and cosmology.  The
isotope is produced in stars which evolve through the
planetary nebula phase.
Planetary nebulae are the final evolutionary phase of low- and
intermediate-mass stars, where the extensive mass lost by the star on
the asymptotic giant branch is ionised by the emerging white
dwarf. This ejecta quickly disperses and merges with the surrounding
ISM. 

The abundance of $^3$He can only be derived from the
hyperfine transition of the ionised $^3$He, which is represented as $^3$He$^+$, these transition can be observed in the radio at the rest frequency of 8.665\,GHz. 
 $^3$He abundances in PNe can
help test models of the chemical evolution of the Galaxy.  

Many hours have been put into trying to detect this line, using telescopes like Effelsberg a 100m dish from the Max Planck Institute for Radio
Astronomy, the National Radio Astronomy Observatory (NRAO)
140-foot telescope, the NRAO Very Large Array, the Arecibo antenna, the Green Bank Telescope, and only just recently, the Deep Space Station 63 antenna from the Madrid Deep Space Communications Complex.

\end{abstract}

\section{Introduction}

Our Universe has been evolving for 13.8\,Gyr. Over these years many
stars formed and ended their lives enriching the
interstellar medium (ISM), and in consequence enriching the Universe
(\cite[Planck Collaboration et al. 2014]{planck14}). Very few elements have been around since the
beginning, formed by the Big Bang nucleosynthesis (BBN).  BBN is
responsible for the formation of most of the helium isotope ($^4$He)
in the Universe, along with small amounts of deuterium (D), the helium
isotope ($^3$He), and a very small amount of the lithium isotope
($^7$Li) (\cite[Alpher, R. A. et al. 1948]{BBN48}).  

The predicted abundance (by number) of $^3$He (relative to H) formed
by the BBN (just after a few minutes after the Big Bang) is
1.0$\times$10$^{-5}$ (\cite[Karakas, A.~I. \& Lattanzio, J.~C. 2014]{karakas2014}). This abundance depends only
on the parameter of the current density of baryonic matter.  The
present interstellar $^3$He abundance, as per all the light elements,
comes from a combination of BBN and stellar nucleosynthesis
(\cite[Wilson, T.~L. \& Rood, R. 1994]{wilson94}). H\,{\sc ii} regions are young objects compared with
the age of the Universe, and represent zero-age objects. Their $^3$He
abundance is the result of 13.8\,Gyr of Galactic chemical
evolution. In between is the Solar System, which traces abundances at
the time of its formation, 4.6\,Gyr ago.

Observed values in pre-solar material (\cite[Geiss, J. 1993]{geiss93}) and the ISM
(\cite[Gloeckler, G. \& Geiss, J. 1996]{gloecker96}) imply that
$^3$He/H$=$(2.4$\pm$0.7)$\times$10$^{-5}$.  These values from the
ISM and pre-solar material are approximately twice of the BBN,
implying that the $^3$He abundance has increased a little in the last
13.8\,Gyr.  On the other hand any hydrogen-burning zone of a star
which is not too hot ($>$7$\times$10$^6$\,K ) will produce $^3$He via
the {\it p-p} chain, implying that stars with masses
$<$2.5\,M$_{\odot}$ are net producers of $^3$He. For these stars, {\it
  p-p} burning is rapid enough to produce D {\it in situ}, and enable
the production of $^3$He (D+ p $\rightarrow$ $^3$He + $\gamma$).
Stellar evolution models indeed predict the formation of $^3$He in
significant amounts by stars of 1--2.5\,M$_{\odot}$, with an abundance
of $^3$He/H $> 10^{-4}$ (\cite[Bania, T.~M. et al. 2010]{bania10}), which would have raised the
current $^3$He abundance to $^3$He/H$\sim5\times10^{-5}$,
substantially higher than observed (\cite[Karakas, A.~I. \& Lattanzio, J.~C. 2014]{karakas2014}).

\indent \cite{galli95} presented ``The $^3$He Problem". According to
standard models of stellar nucleosynthesis there should be a $^3$He/H
abundance gradient in the Galactic Disk and the proto-solar $^3$He/H value should
be less than what is found in the present ISM.  Observations of the  $^3$He
abundance in H\,{\sc ii} regions show almost no enrichment above the
BBN value (\cite[Rood, R.~T. et al. 1979; Bania, T.~M. et al. 2007]{rood79,bania07}). For the $^3$He problem to be solved, the vast majority of
low-mass stars should fail to enrich the ISM.  One suggestion to solve
this problem is by adding extra mixing in the red giant branch (RGB)
stage. This extra mixing adds to the standard first dredge-up to
modify the surface abundances. \cite{eggleton06} estimated that
while 90\% of the $^3$He is destroyed in 1\,M$_{\odot}$ stars, only
40-60\% is destroyed in a 2\,M$_{\odot}$ star model, depending on the
speed of mixing.

\section{Detecting the $^3$He$^+$ emission line}
The abundance of $^3$He can only be derived from the
hyperfine transition at the rest frequency of 8.665\,GHz. Detecting
$^3$He$^+$ in PNe challenges the sensitivity limits of all existing
radio telescopes.  
Composite $^3$He$^+$ average spectrum for six planetary nebulae (NGC 3242, NGC 6543, NGC 6720, NGC 7009,
NGC 7662, and IC 289), using Effelsberg, Arecibo and the Green Bank Telescope observations.
They all consistently show $^3$He$^+$ emission at the $\sim$1\,mK level. Using the same telescopes later on a detection was obtained in NGC 3242 and J320. These detections and the composite spectrum measurements translates into an abundance of $^3$He/H of a few 10$^{-4}$ (\cite[Balser, D.~S. et al. 1997, 1999, 2006]{balser97, Balser1999, Balser2006}). 

Later on more effort was put into this problem, using the Very Large Array, for several tens of hours 3 planetary nebulae were observed (IC\,418, NGC 6572, and NGC 7009), unfortunately no line was detected and only upper limits in the abundance were obtained (\cite[Guzman-Ramirez, L. et al. 2013]{me13}).
Only just recently (beginning of this year) \cite{me16} using the Deep Space Station 63 antenna from the Madrid Deep Space Communications Complex, another $^3$He$^+$ detection was observed in the planetary nebula IC\,418. 
With more than 20 years of looking for these detections, so far we have only 3 objects with more than a 3$\sigma$ detection of $^3$He: J320, NGC 3242, and IC 418. 

\begin{figure}
\centering
\includegraphics[width=7.5cm, height=8.5cm]{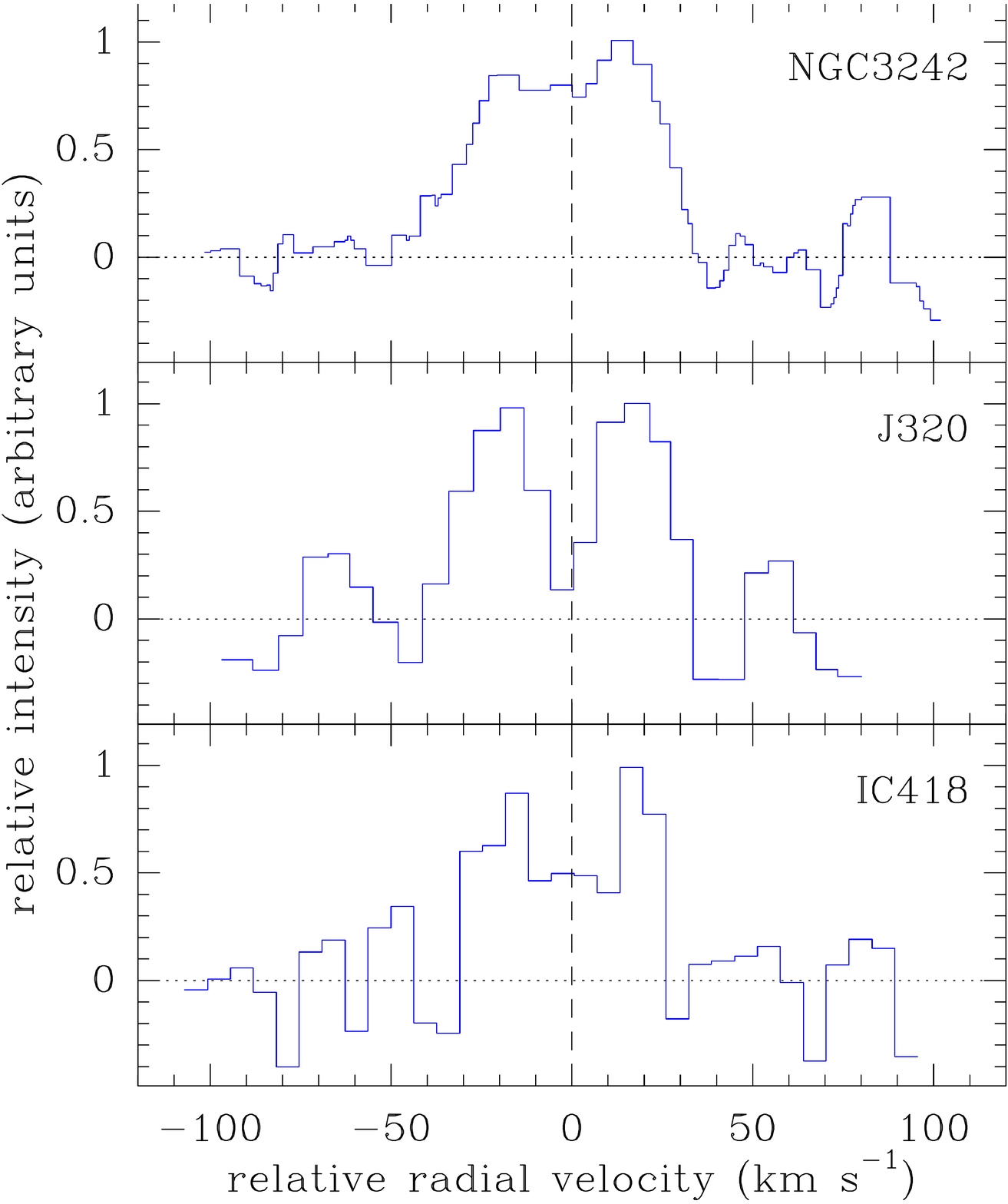}
\caption{All the $^3$He$^+$ line detected in planetary nebulae. The sources are NGC\,3242 
(Balser et al. 1999), J320 (Balser et al. 2006), and IC\,418 (this work). In order 
to facilitate the comparison, the abscissa is presented as radial velocities 
relative the LSR systemic velocity of each source (dashed-line), and the intensity scale as 
fractions of the peak intensity of each source.}
\label{profiles}
\end{figure}   

Two aspects of these detections needs to be considered; Firstly, the derived $^3$He$^+$ abundance is well above model expectations for all the three objects. This tell us that the stars do produce and release its $^3$He into the interstellar medium. 
Secondly, the $^3$He$^+$  line profile for all the three objects shows a double peaked shape. Although the full width of the $^3$He$^+$ line is consistent with the optical expansion velocity, the profile differs from that of the recombination lines, peaking at the outermost velocities.

The two other planetary nebulae have reported $^3$He$^+$ detections: J320
(\cite[Balser, D.~S. et al. 2006]{Balser2006})  and NGC 3242 (\cite[Balser, D.~S. et al. 1997, 1999]{balser97,Balser1999}). Both have double peaked
profiles, similar to IC\,418 (Fig. \ref{profiles}). Both objects have
haloes, in the case of NGC\,3242 possibly as large as 18 by 24\,arcmin
diameter. Whether helium in such a halo could be
photo-ionized by the star is not clear.  
A double-peaked profile could arise from an expanding detached shell,
which is larger than the beam of the telescope. In this case, the outer components of
the profile, which differ from the recombination lines, come from this
large region, whilst the central part of the profile arises from the
inner, ionized nebula. \cite{Balser1999} also proposed a contribution
from a large, low density halo. The emission in the $^3$He$^+$
hyperfine line scales with $\int n\, dr$ (column density), and that of
recombination line with $\int n^2\,dr$. Therefore, a low density but
high mass halo could explain the difference in profiles.

\section{Stellar evolution}
The contribution of PNe to the $^3$He
abundance is crucial for understanding the Galactic chemical evolution. 
The number density of $^3$He$^+$ atoms, $n$($^3$He$^+$), can
be obtained by dividing the column density $N$($^3$He$^+$) by the
averaged optical path, $<\Delta{s}>$, through the source. Representing
a PN as a homogeneous sphere of radius $R=\theta_sD$, where $D$ is the
distance, the optical path at angle from the centre $\theta$ is
$\Delta{s}(\theta)=2\sqrt{R^2-(\theta D)^2}$, and the optical path
averaged over the source becomes $<\Delta{s}>=\pi R/2 = \pi
\theta_sD/2$.  To calculate the fractional $^3$He abundance we 
divide by the H$^+$ density. 

\begin{figure}
\centering
\includegraphics[width=7.5cm, height=8cm]{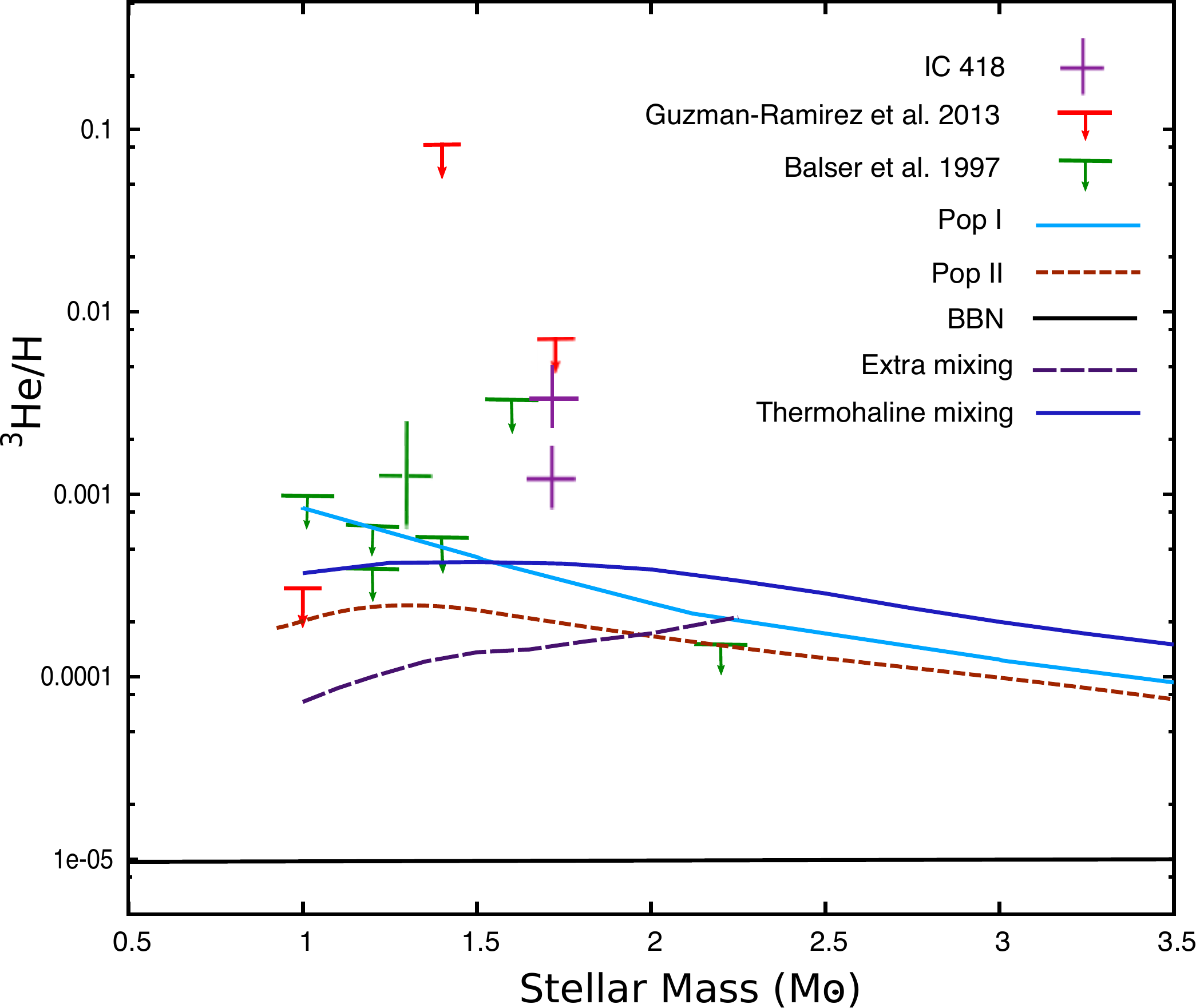}
\caption[$^3$He/H abundance]{Abundances of $^3$He (relative to H) for
  the PN IC\,418 (both purple crosses, for upper and lower estimate),
  the size of the bars correspond to the uncertainties. Red arrows are
  the upper limits from \cite{me13}. The green cross
  represents J320 and the green arrows represent the sample of 6 PNe from
  \cite{balser97}. The curves for Pop I (light-blue line) and II (red
  dashed-line) show the standard abundance of $^3$He taken from
  \cite{weiss96}. The purple dashed-line labelled Extra mixing
  represents the results of models including deep mixing by
  \cite{boothroyd99}, and the blue line shows the stellar models using
  thermohaline mixing (\cite[Charbonnel, C. \& Zahn, J.-P. 2007]{charbonnel07}). The black line is the primordial value of
  $^3$He/H from the BBN (\cite[Karakas, A.~I. \& Lattanzio, J.~C. 2014]{karakas2014}).}
\label{abundances}
\end{figure}

Figure \ref{abundances} shows the $^3$He abundances of the PN IC\,418
(both purple crosses, taken from \cite{me16}).  For comparison the upper limits calculated from
\cite{me13} observations are presented using the red
arrows. \cite{balser97} observations are shown in green: the 
cross represents J320 and the arrows are the upper limits, where
the mass estimates are from \cite{galli97}.  The stellar evolution
models are also presented in the same figure.

\section{Future}

The values observed in these objects proves that $^3$He is produced at the centre of low-mass stars, and is ejected into the interstellar medium at the end of its life. However, the large amounts of $^3$He produced in these models
is at odds with the abundance of $^3$He observed in the interstellar medium and the
Solar System. Further research on other planetary nebulae will be needed to solve the $^3$He problem, but as the emission is extremely weak and hard to detect, many observing hours are needed, or telescopes with higher sensitivity like the Square Kilometre Array will be needed to do follow this research.

%\begin{discussion}
%\end{discussion}

\end{document}